\documentclass{amsart}
\usepackage{wasysym}
\usepackage{amssymb}
\usepackage{amsfonts}
\usepackage{bm}
\usepackage{graphicx}
\usepackage{epsfig}
\usepackage{float}
\usepackage{caption}
\usepackage{tikz}
\usepackage{lipsum}

\newtheorem{thm}{Theorem}[section]
\newtheorem{prop}[thm]{Proposition}
\newtheorem{cor}[thm]{Corollary}
\newtheorem{lemma}[thm]{Lemma}

\newtheorem*{Will}{Corollary (Williams)}
\newtheorem*{Dz}{Corollary (Dziobek)}

\theoremstyle{definition}

\theoremstyle{definition}
\newtheorem{exam}[thm]{Example}
\newtheorem{rem}[thm]{Remark}

\newcommand{\ssp}{\mathbb{W}}

\newcommand{\reals}{\mathbb{R}}

\newcommand{\longvec}{\overrightarrow}

\newcommand{\bary}{\text{bar}}

\numberwithin{equation}{section}

	\title[Moments, Equilibrium Equations and Barycentric Coordinates]{Moments, Equilibrium Equations, and Barycentric Coordinates}
	\author{Eduardo S. G. Leandro\\} 
	
	\address{Universidade Federal de Pernambuco (UFPE) \\
		Departamento de Matem\'atica \\
		Av. Jornalista An\'{\i}bal Fernandes S/N, Recife, PE \\
		50740-560, Brasil\\
		Phone: 55 81 2126-7665, Fax: 55 81 2126-8410.
			}
	\email{eduardo.leandro@ufpe.br}

	\subjclass[2020]{70F10, 37N05, 70Fxx, 37Cxx.}
	\keywords{Affine Geometry, Barycentric Coordinates, Equilibria, $N$-Body Problems, Central Configurations.}
	
\begin{document}
	\sloppy	
	\begin{abstract}
		In a previous article, we applied the classical theory of moments to establish equilibrium equations for rather general finite systems of interacting particles. An equilibrium of such a system occurs if and only if  the corresponding ``zeroth'' moment $\mu_0$ and the first moment $\mu_1$ vanish. Our focus then was on the relationship between $\mu_0$, $\mu_1$ and the second moment $\mu_2$ in order to produce systems of equilibrium equations solely dependent on mutual distances, including  a generalization of the Albouy-Chenciner equations for central configurations in Celestial Mechanics. In the present work, we explore in depth the condition of simultaneously vanishing $\mu_0$ and $\mu_1$ on its own. This condition defines at once a vector space which is studied via barycentric coordinates and elementary exterior algebra, thereby yielding families of equilibrium equations for systems with any consistent number $n$ of particles and dimension $d$.  As before, the present equilibrium equations are homogeneous as well as invariant by oriented isometries (for interactions depending only on mutual distances). Our method also develops and extends the approach by A. Albouy from central configurations of codimension $c=(n-1)-d$ equal to one, the so-called Dziobek configurations, to equilibria of systems of interacting particles of arbitrary codimension. Despite the similar appearance, our generalized Albouy equations constitute new equilibrium equations when $c$ is greater than one, even in the case of central configurations. The paper closes with applications to central configurations, with specific results for $n=5,6$ in dimension $d=2$.
		
	\end{abstract}

	\maketitle
	
	\section{Introduction} \label{Intro}


  Among the most important sets of equilibrium equations in Celestial Mechanics one finds the sets of Dziobek equations~\cite{Dziobek,Hagihara} for central configurations. Such equations have played a relevant role in celebrated results such as the classification of central configurations formed by four points of equal masses by Albouy~\cite{AlbouyII,AlbouyIII}, and the solution of Smale's finiteness problem~\cite{Smale,SmaleII} for four point masses, as well as the solution of the finiteness problem for four point vortices, by Moeckel and Hampton~\cite{Hampton-Moeckel, Hampton-MoeckelII}. 
     
     In his original approach to formulating central configuration equations, Dzio- bek~\cite{Dziobek} employed constrained critical point theory and mutual distances as coordinates. He considered planar configurations formed by an arbitrary number of point masses, and used Cayley-Menger determinants as constraints to guarantee that the configurations had the correct dimension. In order to arrive at his equations, Dziobek noted a remarkable property of the Cayley-Menger determinant, namely, that its derivative with respect to a squared mutual distance can be expressed as a scalar times the product of the oriented areas of a suitably chosen pair of subconfigurations.
     
     Alternative deductions of Dziobek equations by Albouy~\cite{Albouy} and Moeckel~\cite{MoeckelI} did not use this property of Cayley-Menger determinants. Instead,~\cite{Albouy,MoeckelI} followed distinct  algebraic paths. While the approach in~\cite{MoeckelI} still used the typically large Cayley-Menger determinants, the method of~\cite{Albouy} completely avoided such determinants, offering a simpler and rather direct deduction based on barycentric coordinates and elementary linear algebra. Unlike Dziobek, though, both references~\cite{Albouy,MoeckelI} only considered the so-called Dziobek configurations, i.e., configurations of dimension $d=n-2$ formed by $n$ points. In terms of the concept of \emph{codimension}, $c=(n-1)-d$, introduced in~\cite{Leandro}, 
     the Dziobek configurations considered by~\cite{Albouy} and~\cite{MoeckelI} correspond only to the case $c=1$. 
     One of our two main goals is to use the framework provided by the theory of moments~\cite{Leandro} in order to extend Albouy's approach to Dziobek equations from configurations of codimension one to those of arbitrary codimensions. The other main goal is to deduce equilibrium equations directly from the condition of identically vanishing first moment. Rather than remaining confined to Celestial Mechanics, we shall see that our formulations can be applied to a wide range of equilibrium problems in Mechanics.

   This introductory section contains a brief review of the parts of the basic theory of moments which relate to barycentric coordinates; we suggest consulting our previous work~\cite{Leandro} for a more thorough presentation. After the review, we recall the fundamental connection with equilibrium problems for systems of interacting particles possibly subject to certain external forces. Finally, we provide an outline of the remainder of the article.
	
	\subsection{Basic Notions and Results on Moments} A \emph{weighted system} is a pair formed by a finite subset $X$ of an affine space $\mathbb{A}$,  and a function $w$  from $X$ to the field of scalars $K$ of $\mathbb{A}$, that is, $(X,w)$. We refer to $X$ as a \emph{configuration} of points in $\mathbb{A}$, and to $w$ as a \emph{weight function} on $X$.  The scalar $\mu_0$ and  the map $\mu_1$ from $\mathbb{A}$ to the vector space $\longvec{\mathbb{A}}$ of translations of $\mathbb{A}$, which are given by
    \[ \mu_0=\sum_{x \in X} w(x) \quad \text{and} \quad \mu_1(p)= \sum_{x \in X} w(x) \longvec{px}, \ \ \forall p \in \mathbb{A},\]
    are known as the \emph{total weight} (\emph{zeroth moment}) and \emph{first moment} of $(X,w)$, respectively. For Euclidean affine spaces, we can also define the \emph{second moment} $\mu_2:\mathbb{A}\longrightarrow K$ of $(X,w)$ by $\mu_2(p)= \sum_{x \in X} w(x) \longvec{px}^2$, for all $p \in \mathbb{A}$, where the square of a vector stands for the inner product of the vector with itself. The second moment was considered in detail in~\cite{Leandro}, and will not be discussed here; actually, except when required by a physical model, we need not impose the existence of an Euclidean structure on the affine space . It will be convenient and quite fruitful to view $\mu_0$ and $\mu_1$ as linear maps on the weight function $w$. By far, this is the dominant perspective in this article.

	 Consider a fixed weighted system $(X,w)$ in an affine space $\mathbb{A}$.
	
	\begin{lemma} \label{firstmu}
		For all $p,q \in \mathbb{A}$, we have that $ \mu_1(p)-\mu_1(q)=\mu_0\longvec{pq}.$
	\end{lemma}

    \begin{prop} \label{mu1mu0} If $\mu_1$ is the \emph{zero function}, which we write as $\mu_1=\longvec{O}$, then $\mu_0=0$.
    \end{prop}
	
	If $\mu_0 \neq 0$, then we verify that $\mu_1$ is a bijection. The unique point $G \in \mathbb{A}$ such that $\mu_1(G)=\longvec{O}$ 
    is called the \emph{barycenter} of $(X,w)$ and is denoted  by $\bary(X,w)$. 
	
	
	
	\begin{prop}  Let $x \in X$ be such that $w(x) \neq 0$. If $\mu_0=0$, then, for all $p \in \mathbb{A}$, $\mu_1(p)=w(x) \longvec{G_x x}$, where $G_x$ is the barycenter of the weighted system $\left(X\setminus\{x\}, w\big|_{X \setminus \{x\}}\right)$. \label{red1stmom}
	\end{prop}
	
	 The following corollary is an immediate consequence of proposition~\ref{red1stmom}. 
	
	
	\begin{cor} \label{barycentriccoords}
		If $\mu_1=\longvec{O}$, then for every $x \in X$ with $w(x)\neq 0$, we have that $x=\bary\left(X\setminus\{x\}, w\big|_{X \setminus \{x\}}\right)$.
	\end{cor}
	
	A key remark is that, if $X \setminus \{x\}$ in corollary~\ref{barycentriccoords} is a simplex with the same dimension as $X$, then $w\big|_{X \setminus \{x\}}$ defines barycentric coordinates~\cite{Berger} for $x$ with respect to $X \setminus \{x\}$.

	\subsection{Equilibrium Configurations}  \label{Statics} Now suppose $X \subset \mathbb{A}$ is the set formed by the positions of finitely many interacting particles. Assume the force of interaction between the particles at $x,y\in X$ has the usual form~\cite{Arnold}
		\begin{equation} \label{forces}
			\quad	 \longvec{F}_{x,y}=\varphi_{x,y}\longvec{xy}, 
		\end{equation}
		where the coefficients $\varphi_{x,y}$ are scalars and, if $X$ is allowed to vary, the $\varphi_{x,y}$ may be interpreted as functions of the positions of the points in $X$. Newton's third law of motion, or, simply, Newton's third law, corresponds to the symmetry condition $\varphi_{x,y}=\varphi_{y,x}$ .

        We call \emph{equilibrium} a configuration $X$ for which the total force $\longvec{F}_x=\sum_{y \neq x}\longvec{F}_{x,y}$ at $x$ equals $\longvec{O}$ for all $x \in X$. For each $x \in X$, let $(X,w_x)$ be the weighted system defined by
		\begin{equation} \label{weights}
			w_x(y)=\varphi_{x,y}, \ \ \text{if} \ y \neq x, \quad \text{and} \quad w_x(x)=-\sum_{y \neq x } \varphi_{x,y}. 
		\end{equation}
		Each system $(X,w_x)$ has zero total weight and, by a direct calculation, its (constant) first moment equals $\longvec{F}_x$. 
		The next result is a direct consequence of this observation.
	
	\begin{thm} \label{equilibriathm}
		A configuration $X$ of interacting particles subject to forces of interaction of the form~\eqref{forces} is an equilibrium if and only if the corresponding weighted systems $(X,w_x)$, $x \in X$, defined by equations~\eqref{weights} have identically zero first moments.
	\end{thm}
	
	As a matter of fact, the applications of theorem~\ref{equilibriathm} comprise all systems satisfying the ``weaker fundamental hypothesis {\bf(WFH)}" of reference~\cite{Leandro}, namely: \emph{the total interaction \emph{(i.e., the sum of all internal and external forces)} at each point $x$ of the configuration $X$ is a linear combination of the relative position vectors from $x$ to the remaining points of $X$}. Numerous  mechanical systems satisfy the (WFH), so the corresponding equilibrium problems, including the problem of relative equilibria and central configurations in Celestial Mechanics, can be formulated using our procedure. 
    
     From now on, mechanical systems which fulfill hypothesis (WFH) will be called \emph{general}.
    \label{fundamentalhypothesisI} 

    Theorem~\ref{equilibriathm} establishes a fundamental relation 
    between equilibrium configurations of general systems and their moments $\mu_0$ and $\mu_1$. This relation appears in the form of membership in the kernel of the linear map $(\mu_0,\mu_1)$ on the vector space $\ssp(X)$ of the weight functions defined by the interactions~\eqref{weights} between points in the system. The kernel of $(\mu_0,\mu_1)$, denoted by $\ssp_0(X)$, thus plays a chief role in our approach to equilibrium problems, and we shall see that exterior algebra of  $\ssp_0(X)$  constitutes a primary source of equilibrium equations.

\begin{rem}
    A substantial part of the present article is based on our preprint~\cite{LeandroI}. More recently, there appeared the preprints~\cite{AlbouyIV} and~\cite{DiasI} with related contents, albeit based on approaches substantially different from ours.
    In the case of~\cite{DiasI}, the author aims at a generalization of Moeckel's approach to Dziobek determinantal identities. The results in~\cite{DiasI} may suggest possible extensions to some of the formul\ae~ in this paper. In return, our method might simplify the stratification of central configurations by dimension described in~\cite{DiasI}.
\end{rem}

\subsection{Overview of the next sections}
From~\cite{Leandro}, we know that the dimension of $\ssp_0(X)$ equals the codimension $c$ of $X$.
In section~\ref{weightedspaces} we take the next step in the analysis of $\ssp_0(X)$ by providing  
      a method for constructing bases and generating sets for $\ssp_0(X)$. Vectors in the constructed bases and generating sets are expressed in terms of the geometrically meaningful M\"{o}bius volume barycentric coordinates with respect to a simplex of maximal dimension in $X$; we use the term \emph{core} of $X$ in reference to such a simplex.
      Our method 
      is founded on the partition of $X$ into a  core and a cloud of $c$ points. This partition is arranged in a diagram called \emph{Dziobek fan}, which serves as a tool for determining a basis. Proposition~\ref{gensetcard} and remark~\ref{kappa} respectively discuss the cardinality $\kappa=\binom{c+1}{2}$ of some specific generating sets and its connection with Cayley-Menger constraints for mutual distances.


     Section~\ref{extalg} is dedicated to drawing immediate consequences of section~\ref{weightedspaces}. The equality $\dim \ssp_0(X)=c$ readily implies that any collection of $c+1$ weight vectors in $\ssp_0(X)$ is linearly dependent. Theorem~\ref{nice} states this remark in a very brief form using the exterior product, namely  
     \[\bigwedge^{c+1}\ssp_0(X)=\{O\}.\] 
     The above ``codimension -- exterior power formula'' yields a significant number of determinantal identities, and lead to a broad generalization of some formul\ae~obtained by Dziobek for central configurations of codimension one, as exemplified later on by corollary~\ref{nice1}. As illustrated in example~\ref{nongeneric}, these extended Dziobek identities translate into equilibrium equations well suited to exploit geometric properties such as dimension and symmetry.
     Theorem~\ref{basesrep} offers a criterion for membership of a weight function in $\bigwedge^1\ssp_0(X) =\ssp_0(X)$ in the form of a system of degree two homogeneous equations in the values of the weight function and volume barycentric coordinates. In view of theorem~\ref{equilibriathm}, theorems~\ref{nice} and
     ~\ref{basesrep} produce equilibrium equations for all general systems of interacting particles.
     A somewhat surprising connection with Algebraic Geometry arises as we interpret barycentric coordinates as \emph{Plücker coordinates}: Plücker relations~\cite{Harris,Jacobson} for Grassmannians of $d$-dimensional subspaces of $K^{n-1}$ emerge directly from the membership formul\ae ~in theorem~\ref{basesrep}.
     We use the unique Pl\"{u}cker relation for $n=4,d=2$ to exhibit a basis for the first nontrivial instance of the exterior square $\bigwedge^2 \ssp_0(X)$, namely, the case when $X$ is a planar five-point configuration, and apply this basis to equilibrium equations, see lemma~\ref{Williamsgen}, proposition~\ref{eqeqexteriorsquare} and corollary~\ref{generic}. An additional application is made in section~\ref{cc}, where we deduce the formul\ae~by Williams~\cite{Williams} for planar five-body central configurations.
     


       Section~\ref{gde} describes a way of extending the approach by Albouy~\cite{Albouy} to Dziobek theory from configurations of codimension $c=1$ to those with $c>1$. This is achieved by considering matrices whose columns are the weight vectors in the bases constructed in section~\ref{weightedspaces}. The pseudo-inverses of such matrices are used to obtain our \emph{generalized Albouy  equations}~\eqref{gendziobekeq}. Theorem~\ref{gendziobek} applies equations~\eqref{gendziobekeq} as a criterion for symmetric families of weighted systems to have $\mu_1=\longvec{O}$.  Corollary~\ref{dzioriginal} extends the original Dziobek equations for central configurations to general mechanical systems of codimension one, and corollary~\ref{6to15} makes an application of theorem~\ref{gendziobek} to a system of six interacting particles of codimension three. It is noteworthy that our matrix equations~\eqref{gendziobekeq} coincide with Dziobek's equations only for configurations of codimension one (up to sign choices for the volume barycentric coordinates), so they actually constitute new equilibrium equations for general systems of interacting particles. 

       Finally, in section~\ref{cc}, we discuss applications of sections~\ref{extalg} and~\ref{gde} to central configurations of particles of nonzero mass (vorticity, electric charge, etc) interacting under homogeneous potentials with real exponents. We consider both the case of zero and nonzero total mass. Previously known identities~\cite{Dias,Williams} are re-obtained as direct consequences of our previous results. In proposition~\ref{nongen5bp}, we examine the planar five-body central configurations with exactly three collinear bodies, providing some new results, e.g. the symmetry of certain convex, non-strictly convex, configurations. The possibility of writing Dziobek's original equations in matrix form is briefly discussed. 
       We conclude with a simple application of our generalized Albouy equations to prove the impossibility of some classes of planar six-body configurations being central.
     




	\section{Spaces of weight functions with  $\mu_1={\protect\overrightarrow{O}}$}    \label{weightedspaces}
	
	
	As in reference~\cite{Leandro}, section 3, let $X=\{x_1,\ldots,x_n\}$ and let $d=\text{dim}(X)$, be the dimension of (the affine closure of)  $X$. The difference $c$ between the maximum possible value for $d$ and $d$ itself, that is $c=(n-1)-d$, is the \emph{codimension} of $X$; we write $\text{codim}(X)=c$. A weight function $w:X\longrightarrow K$ shall alternatively be represented by the \emph{weight vector} $W \in K^n$ whose coordinates are $w(x_j)=w_j$, $j=1,\ldots,n$. 
    
    After introducing affine coordinates, we define the \emph{(augmented) configuration matrix} of $X$
	\begin{equation} \label{augmented} \mathcal{X}=\left[\begin{array}{ccc}
		1&  \cdots& 1 \\
		x_1& \cdots & x_n
	\end{array}
	\right].
	\end{equation}
	Regard $\mathcal{X}$ as a linear map from $\mathbb{W}(X):=K^n$ to $K^{d+1}$, where $\mathbb{W}(X)$ consists of all possible weight vectors on $X$. We call $\ssp(X)$ the \emph{weight space} of $X$. The kernel of $\mathcal{X}$ is the subspace consisting of weight vectors such that $\mu_0=0$ and $\mu_1=\longvec{O}$. In view of proposition~\ref{mu1mu0}, we simply write
	\[ \text{ker}(\mathcal{X})=:\mathbb{W}_0(X)=\{W \in \mathbb{W}(X) \, \big/ \,  \mu_1=\longvec{O} \}.\]
	The main objective in this section is to describe the space $\ssp_0(X)$. We begin by recalling lemma 5.1 from~\cite{Leandro}.

	
	\begin{lemma} \label{dimnull}
		$\dim\mathbb{W}_0(X)=\text{\emph{codim}}\, (X)$.
	\end{lemma} 

	 So the space $\ssp_0(X)$ for configurations with codimension $c=0$ is the trivial space. For positive codimension, we shall see that a key role is played by configurations with $c=1$. Borrowing the terminology from Celestial Mechanics, such a configuration is termed a \emph{Dziobek configuration}, or \emph{Dziobek subconfiguration}, if it is a proper subset of a larger configuration.

	
	
\begin{figure} 
		\begin{center}
			\scalebox{.4}{\includegraphics{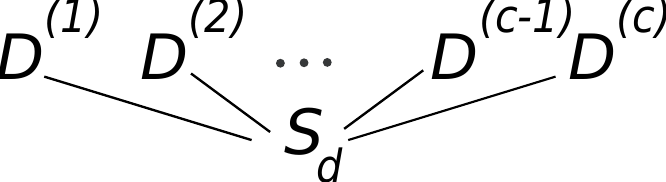}}
			\caption{Sketch of a Dziobek fan. The configuration is the union of the $c$ Dziobek subconfigurations $D^{(1)}, \ldots, D^{(l)}$ containing the core $S_d$. }
			\label{f1}
		\end{center}
	\end{figure}

	Notice that a configuration $X$ has dimension $d$ if and only if $X$ contains a $d$-dimensional simplex $S_d$ and no simplex of higher dimension -- $S_d$ can be constructed from any point in $X$ by successively adding points from $X$ so that, at each step, the dimension increases by one. We refer to any simplex of dimension $d$ contained in $X$ as a \emph{core of} $X$. There are $n-(d+1)=c$ points in $X \setminus S_d$, so $X$ contains exactly $c$ distinct Dziobek subconfigurations $D^{(1)},\ldots,D^{(c)}$ whose intersection is precisely the core $S_d$. The diagram in figure~\ref{f1} depicts one such arrangement, which we call a \emph{Dziobek fan} for $X$.
	
	Consider a Dziobek fan for $X$. If $c=0$, we have that $X$ is the core $S_d$ and, as mentioned before, $\ssp_0(X)$ is the trivial space. The structure of $\ssp_0(X)$ in the cases $c>0$ is explained in the next proposition. For each Dziobek subconfiguration $D^{(l)}$ containing $S_d$, consider the space $\ssp_0(D^{(l)})$, $l=1,\ldots,c,$ and the linear embedding
	\begin{equation} \label{embeddings}
		 \iota_l: \ssp_0(D^{(l)}) \hookrightarrow \ssp_0(X),
   \end{equation}
	which is defined by placing zeros at the coordinates corresponding to the $n-(d+2)=c-1$ points in $X \setminus D^{(l)}$.
	
	\begin{prop} \label{bases}
		To every core $S_d$ of a configuration $X$ of codimension $c>0$ there corresponds a unique direct sum decomposition 
		\[ \ssp_0(X)=\bigoplus_{l=1}^c \iota_l\left(\ssp_0(D^{(l)})\right),\]
		where $D^{(l)}$, $l=1,\ldots,c$, are the Dziobek subconfigurations of $X$ which contain $S_d$.
	\end{prop}
	\begin{proof}
		After possibly reindexing the elements of $X$, we may assume that the core is $S_d=\{x_1,\ldots,x_{d+1}\}$. Let $D^{(l)}=S_d\cup\{x_{d+l+1}\}$, $l=1,\ldots,c$. $D^{(l)}$ has codimension 1, so $\ssp_0(D^{(l)})$ is spanned by a weight vector $(w_1,\ldots,w_{d+1},w_{d+l+1})$ such that
		\begin{equation} \label{barycentric}
			\sum_{j=1}^{d+1} w_j+w_{d+l+1}=0, \qquad \sum_{j=1}^{d+1}w_j x_j+w_{d+l+1}x_{d+l+1}=O.
		\end{equation}
		We must have that $w_{d+l+1} \neq 0$, for all $l=1,\ldots,c$, otherwise the above equations would provide a nontrivial linear combination of the vectors $\longvec{x_1x_j}$, $j=2,\ldots,d+1$, contradicting the hypothesis that $S_d$ is a core.  Using the maps $\iota_l$, see \eqref{embeddings}, we embed each of the spanning vectors for the Dziobek subconfigurations $D^{(l)}$ into $\ssp_0(X)$, thus obtaining the collection of $c$ weight vectors 
		\begin{align*} 
			(*, w_{d+2},0,\ldots,0),\ (*, 0,w_{d+3},\ldots,0),\ \ldots,\ (*,0,\ldots,0,w_n),
		\end{align*}
		where each $*$ represents a (nonzero) vector in $\ssp(S_d)$. This collection is clearly a basis for $\ssp_0(X)$.
	\end{proof}
	
	\begin{rem}   \label{interpretation}
	Equations~\eqref{barycentric} imply that
		\[ (w_1+\ldots+w_{d+1})x_{d+l+1}=w_1x_1+\ldots w_{d+1}x_{d+1}, \]
		with $w_1+\ldots+w_{d+1}\neq 0$, so $w_1,\ldots,w_{d+1}$ are \emph{homogeneous barycentric coordinates} for $x_{d+l+1}$ with respect to $S_d$ (cf. corollary~\ref{barycentriccoords} and comments thereafter).
	\end{rem}
	
\begin{figure} 
		\begin{center} 
			\scalebox{.4}{\includegraphics{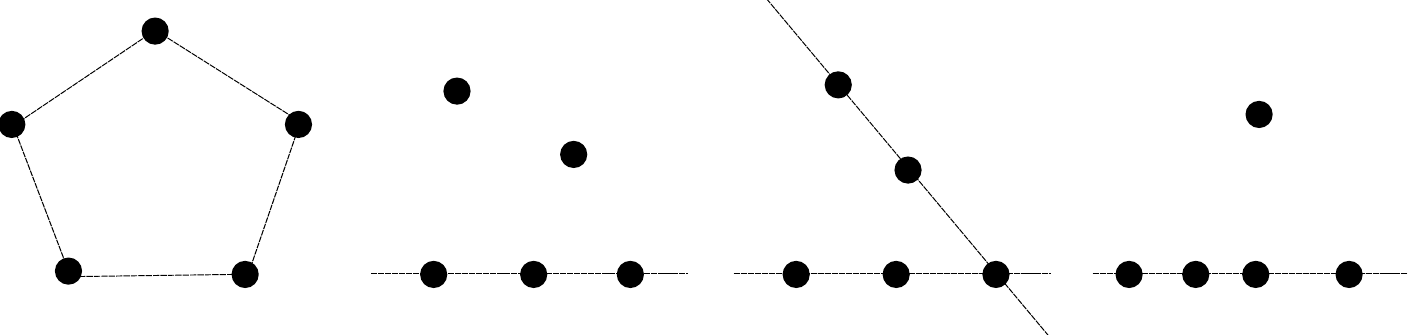}}
			\caption{Planar configurations of five points: generic (left), nongeneric of type $1$ (left center), nongeneric of type 2 (center right), and  nongeneric of type $4$ (right).}
			\label{f2}
		\end{center}
	\end{figure}

	\begin{rem} \label{type} Looking at all possible Dziobek fans for $X$,  proposition~\ref{bases} produces at most $\binom{n}{d+1}=\binom{n}{c}$ bases. The maximum is attained when each subconfiguration of $X$ formed by $d+1$ points is a core of $X$. In this case, we call $X$ a \emph{generic} configuration. A configuration is termed \emph{nongeneric} if it is not generic.  The \emph{type} of a nongeneric configuration refers to the number of its subconfigurations with $d+1$ points that are \emph{not} cores. The possibilities for five points on a plane are depicted in figure~\ref{f2}.
    \end{rem}
	
	For the case $c=1$, M\"{o}bius classical study of barycentric coordinates suggests a geometrically meaningful generator for $\ssp_0(X)$. The following proposition is proved in~\cite{MoeckelI}.
	
	\begin{prop} \label{codim1}
		Let $X=\{x_1,\ldots,x_n\}$, and let $X_j$ be the subconfiguration $X \setminus \{x_j\}$, $j=1,\ldots,n$. Denote by $\mathcal{X}_j$ the augmented configuration matrix of $X_j$.  $X$ has codimension $1$ if and only if $\ssp_0(X)$ is spanned by the weight vector 
		\begin{equation*} 
			W=(-\Delta_1,\Delta_2,-\Delta_3,\ldots,(-1)^n\Delta_n),
		\end{equation*}
		where $\Delta_j=\det(\mathcal{X}_j)$. 
	\end{prop}
	The cases $c>1$ require a definition and additional notation.
	
	The \emph{(oriented) volume} of a configuration $X$ of codimension $c$ is the determinant $\det(\mathcal{X})$ of its augmented configuration matrix $\mathcal{X}$ if $c=0$, and zero otherwise. Let $X_{j_1\ldots j_{\beta}}$ denote the subconfiguration $X\setminus \{x_{j_1},\ldots,x_{j_{\beta}}\}$, where $j_1<\ldots<j_{\beta}$, and let $\mathcal{X}_{j_1\ldots j_{\beta}}$ be the augmented configuration matrix of $X_{j_1\ldots j_{\beta}}$. Denote the volume of $X_{j_1\ldots j_{\beta}}$ by $\Delta_{j_1\ldots j_{\beta}}$. The construction of bases for $\ssp_0(X)$ using the \emph{volume barycentric coordinates} $\Delta_{j_1\ldots j_{\beta}}$ follows the procedure described in the proof of proposition~\ref{bases}. We provide a concrete example.
	
	\begin{exam} \label{spanning1}
		
		Consider seven points $x_1,\ldots,x_7$ forming a two-dimensional configuration $X$. So $X$ has codimension $c=4$. Let us determine a basis for $\ssp_0(X)$ using volumes and proposition~\ref{codim1}. 
		
		Suppose $X_{4567}=\{x_1,x_2,x_3\}$ is a core of $X$; the corresponding volume $\Delta_{4567}$ is, thus, nonzero.  $X_{4567}$ is contained in four Dziobek subconfigurations: $X_{567}$, $X_{467}$, $X_{457}$, $X_{456}$. These Dziobek subconfigurations produce the respective weight vectors 
		\begin{align*}
			W_4 & =(-\Delta_{1567},\Delta_{2567},-\Delta_{3567},\Delta_{4567},0,0,0),\\
			W_5 & =(-\Delta_{1467},\Delta_{2467},-\Delta_{3467},0,\Delta_{4567}, 0,0),\\
			W_6 & =(-\Delta_{1457},\Delta_{2457},-\Delta_{3457},0,0,\Delta_{4567},0),\\
			W_7 & =(-\Delta_{1456},\Delta_{2456},-\Delta_{3456},0,0,0,\Delta_{4567}),
		\end{align*} 
		where the index of each vector is the index of the point added to $X_{4567}$ in order to form the corresponding Dziobek subconfiguration. Since $\Delta_{4567}\neq0$, clearly $W_4,W_5,W_6,W_7$ form a basis of $\ssp_0(X)$.
		
		Now, if $X_{4567}$ is not a core of $X$,  we can replace one of its points with one of the points in $X \setminus X_{4567}$, say replace $x_3$ with $x_4$. If $X_{3567}$ is a core, we consider the weight vectors $W_3'=W_4$ and
		\begin{align*} 
			W_5'&=(-\Delta_{1367},\Delta_{2367},0,-\Delta_{3467},\Delta_{3567},0,0),\\
			W_6'&=(-\Delta_{1357},\Delta_{2357},0,-\Delta_{3457},0,\Delta_{3567},0),\\
			W_7'&=(-\Delta_{1356},\Delta_{2356},0,-\Delta_{3456},0,0,\Delta_{3567}),
  	    \end{align*}
	 which form a basis since $\Delta_{3567}\neq 0$. On the other hand, if $X_{3567}$ is not a core, replace $x_4$ with $x_5$, say, and form the subconfiguration $X_{3467}$. If the latter is a core, the weight vectors $W_3''=W_5$, $W_4''=W_5'$, and
	 \begin{align*}
	 	W_6''&=(-\Delta_{1347},\Delta_{2347},0,0,-\Delta_{3457},\Delta_{3467},0),\\
	 	W_7''&=(-\Delta_{1346},\Delta_{2346},0,0,-\Delta_{3456},0,\Delta_{3467})
	 \end{align*}
	 constitute a basis of $\ssp_0(X)$ for $\Delta_{3467}\neq 0$. If $X_{3467}$ is not a core, then replace $x_5$ with $x_6$, thus obtaining $X_{3457}$. The associated weight vectors $W_3'''=W_6$, $W_4'''=W_6'$, $W_5'''=W_6''$, and
	 \[ W_7'''= (-\Delta_{1345},\Delta_{2345},0,0,0,-\Delta_{3456},\Delta_{3457})\]
	 form a basis as long as $\Delta_{3457}\neq0$. Finally, if $\Delta_{3457}=0$, replace $x_6$ with $x_7$ in $X_{3457}$. The resulting subconfiguration, $X_{3456}$, is necessarily a core, otherwise $X$ is collinear, contradicting the hypothesis $d=2$. The associated weight vectors $W_3''''=W_7$, $W_4''''=W_7'$, $W_5''''=W_7''$, $W_6''''=W_7'''$ do form a basis of $\ssp_0(X)$.  We conclude that the ten-element set
     \begin{equation} \label{spanning}
  \{W_4,W_5,W_6,W_7,W'_5, W'_6,W'_7,W''_6,W''_7, W'''_7\}
     \end{equation}
     spans $\ssp_0(X)$.
	 
	\end{exam} 

\begin{rem} \label{spanning2} It is instructive to observe how, in the planar case, the method in example~\ref{spanning1} takes care of configurations $X$ from generic to nongeneric ones. The first four weight vectors $W_4,W_5,W_6,W_7$ form a basis of $\ssp_0(X)$ for any generic configuration $X$, and for nongeneric configurations we have to make successive substitutions of points and weight vectors. This process ends at a nongeneric configuration of maximal type (see remark~\ref{type}) for which only the minimum $c=4$ of the $\kappa=10$ weight vectors in the corresponding spanning set~\eqref{spanning} do not vanish.
	\end{rem}

     Example~\ref{spanning1} brings to light an interesting general fact.
     
	 \begin{prop} \label{gensetcard}
	 	Any set of $\kappa:=\binom{c+1}{2}$ weight vectors associated with extending a $(d-1)$-dimensional simplex in a $d$-dimensional configuration $X$ to subconfigurations of $X$ with $d+2$ points spans $\ssp_0(X)$.
	 \end{prop}
	 
	 \begin{proof} Notice that $\kappa=\sum_{l=0}^{c}l$. 
     The arguments from  example~\ref{spanning1} naturally generalize to arbitrary $n$ and $d$, with suitable permutations of indices when necessary.
	 \end{proof}
	 
	 \begin{rem} \label{kappa}
	 	  In reference~\cite{Leandro}, we verify that $\kappa=\binom{c+1}{2}$ is precisely the number of constraints that must be imposed on the mutual distances of a configuration of $n$ points to guarantee it has a given codimension $c>0$. By theorem 5.3 of~\cite{Leandro}, the vanishing of $\kappa$ Cayley-Menger determinants provide the necessary constraints. Clearly $\kappa \geq c$, and $\kappa=c$ only when $c=0$, which is the trivial case, or $c=1$. The case $c=1$ is precisely when Dziobek's  approach and the approach in this paper produce essentially the same result. See section~\ref{gde}, corollary~\ref{dzioriginal}, a proof of which can also be found, in the case of central configurations, in Albouy's article~\cite{Albouy}, section 7. 
	 \end{rem}
	
	
	 \section{The Exterior Powers of $\ssp_0(X)$ and Some Applications} \label{extalg}
	 	A basis of $\ssp_0(X)$ produces a basis for each of its exterior powers $ \wedge^l \ssp_0(X)$, $l=1,2,\ldots$~\cite{Greub}.  This elementary fact allows us to write homogeneous polynomial equations involving the entries of vectors in $\ssp_0(X)$ and volume barycentric coordinates. Applications are provided in this section and in  section~\ref{cc}. 

        We begin with a prolific consequence of lemma~\ref{dimnull}.
    

    \begin{thm} \label{nice}
		If $X$ is a configuration of codimension $c$, then $\bigwedge^{c+1}\ssp_0(X)=\{O\}$.
	\end{thm}
	\begin{proof}
		 If a vector space has finite dimension, then its $k$-th exterior power is equal to the trivial space for all $k$ greater than the dimension of the vector space~\cite{Greub}.
	\end{proof}


    Next we provide a characterization of elements of $\wedge^1\ssp_0(X)=\ssp_0(X)$ expressed as a family of homogeneous quadratic identities.
	
	\begin{thm} \label{basesrep}
		Let $X$ be a configuration of dimension $d$ such that (up to relabeling points) $\Delta_{d+2 \ldots n} \neq 0$. We have that $(w_1,\ldots,w_n) \in \ssp_0(X)$ if and only if the identities
		\begin{equation} \label{genthiago}
			(-1)^{i-d}	w_i \Delta_{d+2 \ldots n} =\sum_{l=d+2}^n  w_l \Delta_{i\,d+2 \ldots \widehat{l} \ldots n}, \quad  i=1,\ldots,d+1,
		\end{equation} 
		hold, where $\widehat{l}$ indicates the removal of the index $l$.
	\end{thm}
	\begin{proof}
		By hypothesis, the subconfiguration $\{x_1,\ldots,x_{d+1}\}$ is a core of $X$. Thus we have the following basis of $\ssp_0(X)$
		\begin{align*}
			&(-\Delta_{1d+3\, d+4 \ldots n}, \ldots, (-1)^{d+1}\Delta_{d+1\, d+3\, d+4 \ldots n}, (-1)^{d+2}\Delta_{d+2\, d+3\, d+4 \ldots n},0,\ldots,0),\\
			& \ \ \ \ \ \ \ \ \ \ \ \ \ \ \ \ \ \ \ \ \ \ \ \ \ \ \ \ \ \ \ \ \ \ \ \ \ \ \ \ \ \ \ \ \ \ \ \ \ \ \vdots  \\
			& (-\Delta_{1d+2\, d+3 \ldots n-1}, \ldots, (-1)^{d+1}\Delta_{d+1\, d+2\, d+3 \ldots n-1},0, \ldots,0, (-1)^{d+2}\Delta_{d+2\, d+3 \, d+4\ldots n}).			 
		\end{align*}
		Each of the above weight vectors corresponds to a Dziobek subconfiguration $\{x_1,\ldots,x_{d+1},$ $x_l\}$, with $l=d+2,\ldots,n$ ; we denote the weight vector associated to $l$ by $W_l$.
		
		If $W=(w_1,\ldots,\ldots,w_n) \in \ssp_0(X),$
		we have that $W=\sum_{d+2}^n a_l W_l$, with
		\[a_l=\frac{w_l}{(-1)^{d+2}\Delta_{d+2\ldots n}}, \quad l=d+2,\ldots,n.\] 
		On the other hand, for $i=1,\ldots,d+1$, we have that 
		\begin{equation} \label{rep}
			w_i= \sum_{l=d+2}^n a_l (-1)^i \Delta_{i\, d+2\ldots \widehat{l} \ldots n}=\sum_{l=d+2}^n (-1)^{i-d} w_l \frac{\Delta_{i\, d+2\ldots \widehat{l}\ldots n}}{\Delta_{d+2 \ldots n}},
		\end{equation}
		and the second equality promptly implies equations~\eqref{genthiago}. 
		
		For the converse, given a weight vector $(w_1,\ldots,w_n)$, if we choose $a_l$ so that
		\[ w_l=(-1)^{d+2}a_l \Delta_{d+2 \ldots n}, \ \ \  l=d+2,\ldots,n, \]
		then equations~\eqref{genthiago} imply that
		\[ \Delta_{d+2 \ldots n}\left(w_i-\sum_{l=d+2}^n (-1)^ia_l \Delta_{i\, d+2 \ldots \widehat{l} \ldots n} \right)=0, \ \ \ i=1,\ldots,d+1.\]
		Since $\Delta_{d+2 \ldots n} \neq 0$, we obtain  
		\[ w_i=\sum_{l=d+2}^n (-1)^ia_l \Delta_{i\, d+2 \ldots \widehat{l} \ldots n}, \ \ \ i=1,\ldots,d+1. \]
		Hence $(w_1,\ldots,w_n)=\sum_{d+2}^n a_l W_l$, that is, $W \in \ssp_0(X).$	 
	\end{proof}
	
	\begin{rem}
		The first implication in theorem~\ref{basesrep} does not require
		the hypothesis $\Delta_{d+2 \ldots n} \neq 0$. We illustrate the argument, which can be found in~\cite{Dias}, in the context of central configurations. 
		
		If $(w_1,\ldots,w_n)\in \ssp_0(X)$, then
		\[ \sum_{j \neq k } w_j \longvec{x_kx_j}=\longvec{O}, \quad k=1,\ldots,n.\]
		Set $k=1$ and take the exterior product with
		$\longvec{x_1x_3}\wedge\ldots \wedge \longvec{x_1x_{d+1}}$. Let $\longvec{e}_1,\ldots,\longvec{e}_d$ be the canonical basis of $K^d$. From
		\begin{align*}
			\Delta_{d+2\ldots n} \, \longvec{e}_1 \wedge\ldots \longvec{e}_d&=\longvec{x_1x_2}\wedge \longvec{x_1x_3}\wedge\ldots \wedge \longvec{x_1x_{d+1}}, \ \ \text{and}\\
			\Delta_{2 \, d+2\ldots \widehat{l} \ldots  n} \, \longvec{e}_1 \wedge\ldots \longvec{e}_d&=(-1)^{1-d}\longvec{x_1x_l}\wedge \longvec{x_1x_3}\wedge \ldots \wedge \longvec{x_1x_{d+1}}, \ \ \text{if } \ l>d+1,
		\end{align*}
		we get
		\[(-1)^{2-d} w_2 \Delta_{d+2 \ldots n} =\sum_{l=d+2}^n   w_l \Delta_{2\, d+2 \ldots \widehat{l} \ldots n},\]
		which is~\eqref{genthiago} for $i=2$. 
	\end{rem}
	

\subsection{Application to Pl\"{u}cker Relations} \label{plucker}
 Consider a Dziobek subconfiguration which does not contain the core used in  theorem~\ref{basesrep}. Interestingly, if we write equations~\eqref{genthiago} for the weight vector associated to this Dziobek subconfiguration, we obtain identities corresponding to the Pl\"{u}cker relations used in Algebraic Geometry to describe Grassmannians as projective varieties~\cite{Harris, Jacobson}. 
 
  We examine the case $n=5$, $d=2$ more closely. Notice that the row space of the $2 \times 4$ matrix whose columns are $\longvec{x_1x_2},\ldots,\longvec{x_1x_5}$ has \emph{Pl\"{u}cker coordinates} $p_{12}=\Delta_{45}$, $p_{13}=\Delta_{35}$, $p_{14}=\Delta_{34}$, $p_{23}=\Delta_{25}$, $p_{24}=\Delta_{24}$  and $p_{34}=\Delta_{23}$. This identification between Pl\"{u}cker coordinates and volume barycentric coordinates seems rather general and arises at once by comparing their respective definitions. We have five weight vectors associated to the Dziobek subconfigurations $X_1,\ldots,$ $X_5$ of $X$, namely
			\begin{align*}
				W_1&=(0,-\Delta_{12},\Delta_{13}, -\Delta_{14},\Delta_{15}), \quad W_2=(-\Delta_{12},0,\Delta_{23}, -\Delta_{24},\Delta_{25}),\\
				W_3&=(-\Delta_{13},\Delta_{23}, 0,-\Delta_{34},\Delta_{35}), \quad W_4=(-\Delta_{14},\Delta_{24}, -\Delta_{34},0,\Delta_{45}), 
			\end{align*}
         and $W_5=(-\Delta_{15},\Delta_{25},-\Delta_{35},\Delta_{45},0).$ If $X$ has core $\{x_1,x_2,x_3\}$, then $\Delta_{45}\neq 0$, and $W_4$, $W_5$ form a basis of $\ssp_0(X)$.
		Let us take, for instance, the weight vector associated to the Dziobek subconfiguration $\{x_1,x_2,x_4,$ $x_5\}$, which is $W_3$, and apply formula~\eqref{genthiago} with $i=2$. We find that
		\begin{equation} \label{pluckerdel} \Delta_{23} \Delta_{45}-\Delta_{24}\Delta_{35}+\Delta_{25}\Delta_{34}=0,
        \end{equation}
		which corresponds to the well known unique Pl\"{u}cker relation $p_{12} p_{34}-p_{13}p_{24}+p_{14}p_{23}=0$ defining the Grassmannian $G(2,4)$ of two-dimensional subspaces of $K^4$ as a variety in $K P^5$. We notice that formul\ae~like~\eqref{pluckerdel} can, conversely, be found by applying the Pl\"{u}cker relation for $G(2,4)$ to $2 \times 4$ matrices having columns $\longvec{x_ix_j}$, with $i$ fixed and $j\neq i$.

	\subsection{The space $\bm{\wedge^2 \mathbb{W}_0(X)}$} 
		Next we look at the exterior square of $\ssp_0(X)$. $\wedge^2 \ssp_0(X)$ is the trivial space for simplexes ($c=0$) and Dziobek configurations ($c=1$). We concentrate on the first nontrivial case, namely $c=2$, $n=5$.

\begin{lemma} \label{Williamsgen}
 Let $X$ be a planar configuration formed by five points. The bivector 
 $\sum_{i<j}(-1)^{i+j}\Delta_{ij} E_i\wedge E_j$
	spans $\wedge^2 \ssp_0(X)$, where the vectors $E_1,\ldots,E_5$ form the canonical basis of $\reals^5$.
\end{lemma}
\begin{proof}
    Without loss of generality, we may assume that $\Delta_{45}\neq 0$, so $W_4, W_5$ is a basis of $\ssp_0(X)$. Using the bilinearity and antisymmetry of the wedge product, we obtain
\begin{align*}
    W_4 \wedge W_5 =&(-\Delta_{14}\Delta_{25}+\Delta_{15}\Delta_{24})E_1\wedge E_2+(\Delta_{14}\Delta_{35}-\Delta_{15}\Delta_{34})E_1 \wedge E_3\\
    & -\Delta_{14} \Delta_{45} E_1 \wedge E_4+\Delta_{15} \Delta_{45} E_1 \wedge E_5+(-\Delta_{24}\Delta_{35}+\Delta_{25}\Delta_{34})E_2\wedge E_3\\
    &+\Delta_{45}(\Delta_{24} E_2 \wedge E_4-\Delta_{25} E_2 \wedge E_5-\Delta_{34} E_3\wedge E_4+\Delta_{35} E_3\wedge E_5\\
    &-\Delta_{45} E_4\wedge E_5)=\Delta_{45} \sum_{i<j}(-1)^{i+j}\Delta_{ij} E_i\wedge E_j.
\end{align*}
To verify the last equality, either apply formula~\eqref{genthiago} with $W=W_1$, $i=2,3$, and with $W=W_2$, $i=3$, or use subsection~\ref{plucker}. It follows that the expressions within the three brackets on the left-hand side of the first equality above are $-\Delta_{12} \Delta_{45}$, $\Delta_{13} \Delta_{45}$ and $-\Delta_{23} \Delta_{45}$, respectively. 
\end{proof}


 We apply lemma~\ref{Williamsgen} to equilibrium configurations.

\begin{prop} \label{eqeqexteriorsquare}
Let $y,z$ represent two particles in a general system of five interacting particles, and let $W_y,W_z$ be the weight vectors associated with the weight functions $w_y,w_z$ (see equations~\eqref{weights} for the definition).  If $X$ is a planar equilibrium configuration, then there exists a unique $a_{yz} \in K$ such that, for any pair $i,j=1,\ldots,5$, $i<j$, 
\[ W_y(x_i)W_z(x_j)-W_y(x_j)W_z(x_i)=(-1)^{i+j}a_{yz}\Delta_{ij}.\]
\end{prop}

Clearly, if $X$ is such that $\Delta_{kl} \neq 0$, then the scalar $a_{yz}$ is given by $a_{yz}=(-1)^{k+l}\Delta_{kl}^{-1}$  $[W_y(x_k)W_z(x_l) -W_y(x_l)W_z(x_k)].$ We notice that proposition~\ref{eqeqexteriorsquare} implies the key lemma 2.1 of~\cite{AlbouyIV}.

\begin{cor} \label{generic}
For a planar, general system of five interacting particles such that $X$ is an equilibrium configuration with core $\{x_1,x_2,x_3\}$, the weight vectors $W_x$ and $W_y$ are collinear if and only if $W_x(x_5)W_y(x_4)=W_x(x_4)W_y(x_5)$.
\end{cor}
\begin{proof}
    In terms of the exterior product, collinearity  of $W_x$ and $W_y$ means $W_x \wedge W_y=0$. The hypotheses imply $\Delta_{45} \neq 0$ and, from proposition~\ref{eqeqexteriorsquare}, it follows that $a_{xy}=0$ if and only if $W_x(x_5)W_y(x_4)=W_x(x_4)W_y(x_5)$. But $a_{xy}=0$ is equivalent to $W_x $ $\wedge W_y=0$.
\end{proof}

			\section{Generalized Albouy Equations} \label{gde}
		
		Theorem~\ref{equilibriathm} motivates the  formulation of equations simultaneously satisfied by families $\mathfrak{F}=\{ (X,f_j)\}_{j=1}^n$ of weighted systems with configuration $X$. We use the notation of section~\ref{weightedspaces}. Let us define the \emph{weight matrix} of $\mathfrak{F}$ as the $n\times n$ matrix $\mathcal{F}$ whose entries are $\mathcal{F}_{ij}=f_j(x_i)$, $i,j=1,\ldots,n$. If $\mathcal{F}$ is a symmetric matrix, we refer to the family $\mathfrak{F}$ as a \emph{symmetric} family of weighted systems. Examples of such families are the systems of interacting particles $\{(X,w_j)_{j=1}^n\}$ that obey Newton's third law, see the first paragraph of subsection~\ref{Statics}. 	
		
		Suppose each $(X,f_j) \in \mathfrak{F}$ has $\mu_1=\longvec{O}$. If $W_1,\ldots,W_c$ is a basis of $\ssp_0(X)$, there exists a $c \times n$ matrix  $\mathcal{A}$ with entries in $K$ such that
		\begin{equation} \label{dziobek-1}
			\mathcal{F}=\mathcal{W}\mathcal{A},
		\end{equation}
		where $\mathcal{W}$ is the matrix whose columns are the basis vectors; such a matrix will be referred to as a \emph{basis matrix} of $\ssp_0(X)$. If $\mathcal{W}^T$ admits a right inverse -- for instance, if $K=\mathbb{R},\mathbb{C}$, we can take $\mathcal{W}(\mathcal{W}^T\mathcal{W})^{-1}$~\cite{Penrose} -- then there exists a $c \times c$ matrix $\Sigma$ such that $\mathcal{A}=\Sigma\mathcal{W}^T$. Furthermore, using a left inverse for $\mathcal{W}$,  we verify that $\mathfrak{F}$ is a symmetric family  if and only if $\Sigma$ is symmetric, and, in this case, we can rewrite equation~\eqref{dziobek-1} as
		\begin{equation} \label{gendziobekeq}	\mathcal{F}=\mathcal{W}\Sigma \mathcal{W}^T.
		\end{equation}
		
		\noindent Evidently, by setting $\Sigma \mathcal{W}^T=\mathcal{A}$ in equation~\eqref{gendziobekeq}, we recover equation~\eqref{dziobek-1}. Thus the following statement has been proved.
		
		\begin{thm} \label{gendziobek}
			Let the field of scalars be either $\reals$ or $\mathbb{C}$. The weight vectors of a symmetric family $\mathfrak{F}=\{ (X,f_j)\}_{j=1}^n$ of weighted systems with $X=\{x_1,\ldots,x_n\}$ are in $\ssp_0(X)$ if and only if equation~\eqref{gendziobekeq} is satisfied by the weight matrix $\mathcal{F}=(f_j(x_i))_{1\leq i,j\leq n}$ of $\mathfrak{F}$, for some $n\times c$ basis matrix $\mathcal{W}$ of $\ssp_0(X)$ and some $c \times c$ symmetric matrix $\Sigma$.
		\end{thm}
		
		 We shall discuss applications to $n$-body central configurations in the next section.  Remarks 3.4 and 3.5 of~\cite{Leandro} contain possible additional applications of theorem~\ref{gendziobek}. 
         
          To conclude this section, we obtain Dziobek equations for general systems of interacting particles having $c=1$, and derive a result for configurations with $n=6$, $c=3$.

         \begin{cor} \label{dzioriginal}
             Let $K=\mathbb{R}$ and suppose $\{(X,w_j)\}_{j=1}^n$ is a system of interacting particles obeying Newton's third law. Let $W$ be the weight vector in proposition~\ref{codim1}. $X$ is an equilibrium configuration of codimension one if and only if $W\neq 0$ and, for some $\sigma \in \mathbb{R}$,
             \[ w_j(x_i)= (-1)^{i+j}\sigma\Delta_i\Delta_j, \quad 1 \leq i<j\leq n.\]     
         \end{cor}

         \begin{cor}\label{6to15}
           For a symmetric family of weighted systems $\mathfrak{F}=\{ (X,f_j)\}_{j=1}^6$ on a planar six-point configuration $X$ such that $f_j(x_i)=0$ for the six pairs $\{i,j\}=\{1,4\},\{2,5\}, \{3,6\}, \{4,5\},\{4,6\},\{5,6\}$,  and $\{x_i,x_j,x_k\}\subset X$ form a triangle for the four triples $\{i,j,k\}=\{1,2,3\},\{1,2,6\},\{1,3,5\},\{2,3,4\}$, we have that all $f_j$ vanish identically.
         \end{cor}
		\begin{proof}
		    Consider the basis of $\ssp_0(X)$ formed by $(-\Delta_{156}, \Delta_{256},-\Delta_{356},\Delta_{456},0,0)$, $(-\Delta_{146}, \Delta_{246},-\Delta_{346},0,\Delta_{456},0)$ and $(-\Delta_{145}, \Delta_{245},-\Delta_{345},0,0,\Delta_{456})$. Let $\mathcal{W}$ be the corresponding basis matrix. Equation~\eqref{gendziobekeq}  imply  
            \[ f_{45}=\sigma_{12}\Delta_{456}^2, \quad f_{46}=\sigma_{13}\Delta_{456}^2, \quad \text{and} \ \ f_{56}=\sigma_{23}\Delta_{456}^2.\]
            By hypothesis, the left-hand sides of the above equations equal zero, so $\sigma_{12}=\sigma_{13}=\sigma_{23}=0$, as long as $\Delta_{456}\neq 0$. Substituting these in equation~\eqref{gendziobekeq}, we obtain
            \[ f_{14}=-\sigma_{11}\Delta_{156}\Delta_{456}, \quad f_{25}=\sigma_{22}\Delta_{246}\Delta_{456}, \quad \text{and } \ \ f_{36}=-\sigma_{33}\Delta_{345}\Delta_{456}.\]
            Again by hypothesis, the left-hand sides vanish and, since $\Delta_{156}\neq 0$, $\Delta_{246} \neq 0$ and $\Delta_{345}\neq 0$, we have that $\sigma_{11}=\sigma_{22}=\sigma_{33}=0$. Therefore $\Sigma$ is the zero matrix, and equation~\eqref{gendziobekeq} imply $\mathcal{F}=0$, which is equivalent to the desired conclusion. 
		\end{proof}

		\section{Some Applications to central configurations} \label{cc}

         This section complements section 6 of~\cite{Leandro}, where further information and results on central configurations can be found.
		
		A \emph{configuration of bodies (particles, point masses) in} $\mathbb{E}^N$ is a weighted system $(X,m)$ where $m$ (the ``mass") is a nonvanishing function. 
        The interactions between the bodies are such that $j^{\text{th}}$ body has acceleration
		\begin{equation*} 
			\longvec{\gamma}_j = \sum_{i\neq j} m_i s_{ij}^{a} \longvec{x_jx_i}, \quad  s_{ij}=\longvec{x_ix_j}^2, \ \ a\in \mathbb{R},
		\end{equation*}
		 The pair $(X,m)$ is a \emph{central configuration} if there exist a vector $\longvec{\gamma}_{O}$, a point $x_{O}$ and a real number $\lambda$ such that 
		\begin{equation} \label{cceq}
			\longvec{\gamma}_j-\longvec{\gamma}_O=\lambda \longvec{x_Ox_j}, \quad j=1,\ldots,n.
		\end{equation}
		Albouy~\cite{Albouy} has shown that, if $a=0$, then any configuration is central, and if $a \neq 0$ and $d=n-1$, i.e, if $c=0$, then $X$ is a regular simplex for all values of the \emph{total mass} $\mu_0$. Henceforth we suppose that $a \neq 0$ and, as in~\cite{Albouy,Leandro}, tacitly assume that $\lambda/\mu_0<0$ whenever $\mu_0 \neq 0$.
		
		
		
		
		\begin{prop} \label{zeromuzero}
			If $(X,m)$ is a central configuration with $\mu_0=0$, then $X$ has dimension $d\leq n-2$, and if $\Delta_{d+2 \ldots n} \neq 0$, then
				\begin{equation*} 
					(-1)^{i-d}	m_i \Delta_{d+2 \ldots n} =\sum_{l=d+2}^n  m_l \Delta_{i\,d+2 \ldots \widehat{l} \ldots n}, \quad  i=1,\ldots,d+1.
				\end{equation*} 
		\end{prop}
		\begin{proof} As noted in~\cite{Leandro}, equations~\eqref{cceq} imply that $\lambda \mu_1(x_O)=-\mu_0 \longvec{\gamma}_O.$ So $m \in \ssp_0(X)$, and the assertions are immediate  consequences of lemma~\ref{dimnull}  and theorem~\ref{basesrep}.
		\end{proof}

		From now on we suppose that $\mu_0 \neq 0$. As proved in~\cite{Albouy}, $(X,m)$ is a central configuration if and only if $\longvec{\gamma}_O$ is the null vector and $x_O=\bary(X,m)$ is the \emph{center of mass} of $(X,m)$. The central configuration equations can thus be put in the form 
		\begin{equation} \label{cceqalt}
			\sum_{i \neq j} m_i S_{ij} \longvec{x_jx_i}=\longvec{O}, \quad j=1, \ldots,n,
		\end{equation}
		where $S_{ij}=s_{ij}^{a}+\frac{\lambda}{\mu_0}$. Equations~\eqref{cceqalt} imply that, for each $j$, a weighted system $(X,C_j)$ with $\mu_1=\longvec{O}$ can be defined, namely
		\begin{equation} \label{ccweight}
			C_j(x_i)=m_i S_{ij}, \quad j \neq i,\ \  C_j(x_j)=-\sum_{i \neq j} m_iS_{ij},\ \ \text{for} \ \ j=1,\ldots,n.
		\end{equation}
		For convenience, we write $C_j(x_j)=m_j S_{jj}$. We have just proved the next theorem.
		
		\begin{thm} \label{cczeroweights}
			A configuration of particles $(X,m)$ with $\mu_0 \neq 0$ is a central configuration if and only if each weighted systems $(X,C_j)$, with $C_j$ defined by equations~\eqref{ccweight}, $j=1,\ldots,n$, have first moment equal to the zero function.
		\end{thm}
		Therefore we can apply the theory developed in the preceding sections to the study of central configurations with $\mu_0 \neq 0$. In particular, we recover and extend some known results. 


        \begin{cor}  \label{nice1}
			Let $(X,m)$ be a central configuration with $\mu_0 \neq 0$ and codimension $c$. Every $(c+1)\times (c+1)$-minor of any $n \times (k+l)$-matrix, with $k+l=c+1$, formed by $k$ central configuration weight vectors $C_j$ (see equations~\eqref{ccweight}) and any $l$ weight vectors in $\ssp_0(X)$ (for instance, those corresponding to Dziobek subconfigurations of $X$, see section~\ref{weightedspaces}) vanishes. In particular, any $(c+1) \times (c+1)$-minor of the $n \times n$ matrix formed by the $S_{ij}$ is zero.
		\end{cor}
		\noindent Corollary~\ref{nice1} follows immediately from theorems~\ref{nice},~\ref{cczeroweights} and the fact that $m$ is nonvanishing. It extends corollary 6.4 of~\cite{Leandro}.

        The equations in the next corollary are a direct consequence of theorems~\ref{basesrep} and~\ref{cczeroweights} and correspond to the formul\ae~ in theorem 2.2 of~\cite{Dias}. Notice, however, that formul\ae~ in~\cite{Dias} do not form a complete set of central configuration equations.
		
		
		\begin{cor} \label{Diaseqs}
			$(X,m)$ is a central configuration  with $\mu_0 \neq 0$, of dimension $d\leq n-2$ and such that $\Delta_{d+2 \ldots n} \neq 0$ if and only if
			\begin{equation*} 
				(-1)^{i-d}	m_i S_{ij}\Delta_{d+2 \ldots n} =\sum_{l=d+2}^n  m_l S_{lj} \Delta_{i\,d+2 \ldots \widehat{l} \ldots n}, \quad  i=1,\ldots,d+1, \ j=1,\ldots,n. 
			\end{equation*} 
		\end{cor}

         Newt we focus on central configurations of the planar five-body problem.
	
	
     Since the central configuration weight vectors $C_k$ and $C_l$ are both in $\ssp_0(X)$, proposition~\ref{eqeqexteriorsquare} states that, for some $a_{kl} \in \reals$,
	\[C_k(x_i)C_l(x_j)-C_k(x_j)C_l(x_i)=m_im_j(S_{ik}S_{jl}-S_{il}S_{jk})= (-1)^{i+j}a_{kl}\Delta_{ij}, \]
	with $i,j,k,l=1,\ldots,5$, $i<j$. By eliminating the scalars $a_{kl}$ from the above equations, as we did in the proof of theorem~\ref{basesrep}, we obtain equations due to Williams~\cite{Williams} (reference~\cite{Chen-Hsiao} contains a clear deduction). 

    \begin{Will} 
     If $X$ is a five-body central configuration then, given indices $i_0,i_1,j,k,l=1,2,3,4,5$, with $i_0,i_1<j$, we have that
     \[(-1)^{i_1} m_{i_0}(S_{i_0k}S_{jl}-S_{jk}S_{i_0l})\Delta_{i_1j}=(-1)^{i_0}m_{i_1}(S_{i_1k}S_{jl}-S_{jk}S_{i_1l})\Delta_{i_0j}.\]
    \end{Will}
		
		The next example illustrates the use of determinants provided by corollary~\ref{nice1} to obtain equations tailored for particular families of central configurations.
		\sloppy
		\begin{exam} \label{nongeneric}
			Consider a planar configuration $X=\{x_1,x_2,x_3,x_4,x_5\}$ and let $x_ix_j$ denote the line through the points $x_i$ and $x_j$. Suppose $x_3 \in x_1x_2$ and $x_4,x_5 \notin x_1x_2$. In the terminology of section~\ref{weightedspaces}, $X$ is a nongeneric configuration of type 1 (see figure~\ref{f2}, page \pageref{f2}).

			Corollary~\ref{nice1} imply that, for $i,j,k=1,2,3$ and $l=4,5$, we have that
			\begin{equation} \label{determinants}
				\left|\begin{array}{ccc}
					m_3 S_{3j}&*&-\Delta_{3l}\\
					m_4 S_{4j}&-\Delta_{i4}&0\\
					m_5 S_{5j}&\Delta_{i5}&0
				\end{array}
				\right|=0 \ \ \text{and} \ \
				\left|\begin{array}{ccc}
					m_i S_{ij}&*&(-1)^i\Delta_{il}\\
					m_k S_{kj}&*&(-1)^k\Delta_{kl}\\
					m_4 S_{4j}&-\Delta_{i4}&0
				\end{array}
				\right|=0,
			\end{equation}
			for any central configuration $(X,m)$ with $\mu_0\neq 0$.  We have the following sets of necessary conditions for such $(X,m)$.

			\begin{prop} \label{nongen5bp}
				Suppose $(X,m)$ is a planar central configuration of five particles with nonzero total mass $\mu_0$, and such that $x_3 \in x_1x_2$ and $x_4,x_5 \notin x_1x_2$. Let $h_l$ be the distance from $x_l$ to $x_1x_2$.
				\begin{itemize}
					\item [(a)] For every $j=1,2,3,$ we have that
					\[m_4 S_{j4}h_4=\pm m_5 S_{j5}h_5,\]
					where the $+$ (resp. $-$) sign occurs if $x_4$ and $x_5$ are on opposite sides (resp. the same side) of $x_1x_2$. 
					\item[(b)] Suppose $x_4$ and $x_5$ are on opposite sides of $x_1x_2$. If $h_4=h_5$, and all the masses have the same sign, then $x_1x_2$ is the perpendicular bisector of $x_4x_5$, and $m_4=m_5$. 
					\item[(c)] Suppose $x_4$ and $x_5$ are on the same side of $x_1x_2$.
						If $s_{24}^a$ or $s_{25}^a$ equal $\lambda/\mu_0$, then $s_{24}^a=s_{25}^a=\lambda/\mu_0$, and
						\begin{equation*} \label{cool}
							m_1 S_{12}s_{12}^{1/2}= m_3 S_{23}s_{23}^{1/2}.
						\end{equation*}
						In particular, for $a<-1/2$ and $m_1=m_3$, particle $x_2$ is at the midpoint of $x_1$ and $x_3$. Moreover, if the lines $x_1x_2$ and $x_4x_5$ are parallel, then, for all $m_2$, the configuration is an isosceles trapezoid and $m_4= m_5$.
				\end{itemize}
			\end{prop}
			\begin{proof}
				Item (a) follows from the first determinant in equations~\eqref{determinants} and from $\frac{\Delta_{j4}}{\Delta_{j5}}=\pm \frac{h_5}{h_4}$. For (b), notice $h_4=h_5$ forces $x_1x_2$ to intersect $x_4x_5$ at the midpoint of $x_4$ and $x_5$. Apply the Perpendicular Bisector Theorem~\cite{Albouy} and item (a) to draw the stated conclusions. To obtain the equation in (c), we substitute $S_{24}=0$ in the second determinant of~\eqref{determinants} with $i=1,j=2,k=3,l=4$ and use the equation $\frac{\Delta_{14}}{\Delta_{34}}= \frac{s_{23}^{1/2}}{s_{12}^{1/2}}$.  The remaining statements in (c) follow from the fact that $f(x)=x^{2a+1}-(\lambda/\mu_0) x$, $x>0$, is an injective function for any $a<-1/2$, and from item (a). Indeed, $h_4=h_5$ by hypothesis, so $X$ is an isosceles trapezoid and, by symmetry, $S_{14}=S_{35}$ and $S_{34}=S_{15}$. Thus, by (a),
				\[ m_5 S_{15}=-m_4 S_{14}=-m_4 S_{35}=m_4 \left(-\frac{m_4}{m_5} S_{34}\right) \ \Longrightarrow \ m_4=\pm m_5.\]
				But $m_4=-m_5$ implies $S_{14}=S_{15}$, which is not possible.
			\end{proof}
			
			References~\cite{Chen-Hsiao,Gidea-Llibre} describe a family of planar central configurations of five bodies under the Newtonian gravitational potential (i.e., for $a=-3/2$) which fulfills the conditions in item (c) of proposition~\ref{nongen5bp}. In both~\cite{Chen-Hsiao} and~\cite{Gidea-Llibre}, the masses are positive and the symmetry of $(X,m)$ is assumed from start. On the other hand, the hypothesis $s_{24}^a=\lambda/\mu_0$ is part of the conclusions in~\cite{Chen-Hsiao}, which also discusses the changes in the shape of the isosceles trapezoid as $m_2$ changes. References~\cite{Cornelio,Gidea-Llibre} examine configurations considered in item (b) of proposition~\ref{nongen5bp}. 
		\end{exam}
		
 Lastly we apply the corollaries of theorem~\ref{gendziobek} to central configurations. We must consider the family $\mathfrak{F}=\{(X,m_j C_j)\}_{j=1}^n$, so $\mathcal{F}=\mathcal{C}\mathcal{M}$, where $\mathcal{C}$ is the matrix whose $j^{\text{th}}$ column is the weight vector $C_j$ defined by~\eqref{ccweight}, and $\mathcal{M}$ is the diagonal matrix with entries $m_1,\ldots,m_n$ on the main diagonal.  From corollary~\ref{dzioriginal} we obtain the following familiar statement.


\begin{Dz}
     $(X,m)$ is a central configuration with $\mu_0\neq 0$ and codimension one if and only if the vector $W$ in proposition~\ref{codim1} is nonzero, and for some $\sigma \in \mathbb{R},$ there hold the identities
     \begin{equation} \label{dziobekclassic} m_i m_j S_{ij}= (-1)^{i+j}\sigma  \Delta_i \Delta_j, \quad 1\leq i<j \leq n.
\end{equation}
\end{Dz}   


The above are, up to sign convention for the volume barycentric coordinates, the classical Dziobek equations for central configurations of codimension one~\cite{Albouy,Dziobek,MoeckelI}. 

\begin{rem}
	 It must be stressed  that equations~\eqref{gendziobekeq} \emph{do not coincide} with Dziobek's equations for codimensions $c>1$. Nevertheless, the special form of equations~\eqref{dziobekclassic} can be directly extended to codimensions greater than one by arbitrarily selecting $\kappa=\binom{c+1}{2}$ subconfigurations of $d+2$ points, constructing the respective weight vectors using proposition~\ref{codim1} and embeddings similar to the $\iota_l$, see~\eqref{embeddings}, and forming the corresponding $n \times \kappa$ matrix $\mathcal{W}$. 
     The $c \times c$ symmetric matrix $\Sigma$ in~\eqref{gendziobekeq} is replaced by a $\kappa \times \kappa$ diagonal matrix $\mathcal{L}$, so instead of~\eqref{gendziobekeq}, we would write the equation
	 \begin{equation} \label{dziobekgen}
	 	\mathcal{F}=\mathcal{W} \mathcal{L}\mathcal{W}^T.
	 \end{equation}
	  As previously explained, Dziobek used constrained critical point theory to obtain equations of the form~\eqref{dziobekgen}. He argued that the derivative of a Cayley-Menger determinant with respect to a squared mutual distance can be expressed simply as a constant times the product of a pair of oriented volumes of suitable subconfigurations~\cite{Dziobek, Hagihara}.	
      In the context of critical point theory, the $\kappa$ entries on the main diagonal of $\mathcal{L}$ must be viewed as Lagrange multipliers. 
		Interestingly, the number of parameters defining the symmetric matrix $\Sigma$ in the generalized Albouy equations~\eqref{gendziobekeq} equals the number $\kappa$ of Lagrange multipliers in Dziobek's approach.   However, an explanation for this fact and a deduction of equation~\eqref{dziobekgen} by our methods remain to be provided.
	\end{rem}

    \begin{cor} \label{6to15cc}
        There do not exist any planar central configuration of the six-body problem such that the six distances $r_{14}, r_{45},r_{46}, r_{25}, r_{56}$ and $ r_{36}$ are equal to $(-\lambda/\mu_0)^{\frac{1}{2a}}$,  and the oriented areas $\Delta_{156},\Delta_{246},\Delta_{345},$ and $\Delta_{456}$ are nonzero.
    \end{cor}
    \begin{proof}
        From corollary~\ref{6to15}, it follows that all mutual distances must be equal to $(-\lambda/\mu_0)^{\frac{1}{2a}}$. So the configuration is a regular simplex, and therefore cannot be planar.
    \end{proof}

		\begin{sloppypar}
		\noindent {\bf Acknowledgments.} I would like to express my gratitude to the participants of the Mathematical Physics Seminar in Recife, as well as those of events organized by Luis Fernando Mello, Antonio Carlos Fernandes, and Lei Zhao, for their comments and suggestions. I also thank Thiago Dias for clarifying some key parts of his work~\cite{DiasI}.\\
        \end{sloppypar}
		
		\noindent {\bf Data availability statement.} The author declares that the data supporting the findings of this study are available within the paper.\\
		
		\noindent {\bf Funding/Competing interests statement.} The author has no relevant financial or non-financial interests to disclose. No support was received from any organization for the submitted work.



\begin{thebibliography}{99}
			
			\bibitem{AlbouyII} Albouy, A.: Sym\'etrie des configurations centrales de quatre corps. C. R. Math. Acad. Sci. Paris 320, S\'erie I, 217-220 (1995).
			
			\bibitem{AlbouyIII} Albouy, A.: The symmetric central configurations of four equal masses. Contemp.
			Math. 198, 131--135 (1996).
			
			
			
			\bibitem{Albouy} Albouy, A.: On a Paper of Moeckel on Central Configurations. Regul. Chaotic Dyn. Vol. 8, No.2, 133--142 (2002).        
			

            \bibitem{AlbouyIV} Albouy, A., Sun, J.:  On the eigenvalues of a central configuration, arXiv:2512.00661 [math-ph], https://doi.org/10.48550/arXiv.2512.00661 (2025)

			
			\bibitem{Arnold} Arnold, V. I.: Mathematical Methods of Classical Mechanics, Second edition. Springer-Verlag, New York, NY (1989).
			
			
			\bibitem{Berger} Berger, M.: G\'eom\'etrie 1. Cassini, Paris (2016).
			
			
			
			
			\bibitem{Chen-Hsiao} Chen, K.-C., Hsiao, J.-S.: Convex central configurations of the $n$-body problem which are not strictly convex.
			J. Dynam. Diff. Equations 24, 119--128 (2012).

            \bibitem{Cornelio} Cornelio, J. L., Álvarez–Ramírez, M., Cors, J. M.: A family of stacked central configurations in the planar five-body problem. Celestial Mech. Dyn. Astr. 129, 321--328 (2017). 
			
			
			\bibitem{Dias} Dias, T.:  New equations for central configurations and generic finiteness. Proc. Amer. Math. Soc. 145, Number 7, 3069--3084 (2017).


\bibitem{DiasI} Dias, T.: Brehm-Wintner-Conley Dimension, Plücker Coordinates, and Generalized Dziobek-Williams Equations for Central Configurations, arXiv:2608.07771 [math.DS], https://doi.org\\
/10.48550/arXiv.2608.07771 (2026).

            
			
			
			\bibitem{Dziobek} Dziobek, O.: Ueber einen merkw\"{u}rdigen Fall des Vielk\"{o}rperproblems. Astron. Nach. 152, 33--46 (1900).
			
			
			\bibitem{Gidea-Llibre} Gidea, M., Llibre, J.: Symmetric planar central configurations of five bodies: Euler plus two. Celestial Mech. Dynam. Astronom. 106, 89--107 (2010).
			
			
			
			\bibitem{Greub} Greub, W. H.: Multilinear Algebra. Springer-Verlag, Berlin$\cdot$Heidelberg (1967).
			
			\bibitem{Hagihara}  Hagihara, Y.: Celestial Mechanics I. MIT Press, Cambridge, MA (1970).
			
			\bibitem{Hampton-Moeckel} Hampton, M., Moeckel, R.: Finiteness of relative equilibria of the four-body problem. Invent. Math. 163, 289--312 (2006).
			
			\bibitem{Hampton-MoeckelII} Hampton, M., Moeckel, R.:
			Finiteness of stationary configurations of the four-vortex problem.
			Trans. Am. Math. Soc. 361, 1317--1332 (2009).
			
			\bibitem{Harris} Harris, J.: Algebraic Geometry: A First Course. Springer-Verlag, New York, NY (1992).
			
			\bibitem{Jacobson} Jacobson, N.: Finite Dimensional Division Algebras over Fields. Springer-Verlag, Berlin$\cdot$Hei- delberg (1996), Corrected 2nd printing (2010).
			
			

            \bibitem{Leandro} Leandro, E. S. G.: Moments, Equilibrium Equations, and Mutual Distances. Z. Angew. Math. Phys. 77, 228 (2026). 

            \bibitem{LeandroI} Leandro, E. S. G.: Moments and the Basic Equilibrium Equations, arXiv:2507.13865v3 [math-ph],
https://doi.org/10.48550/arXiv.2507.13865 (2025).
			
			
			
			\bibitem{MoeckelI} Moeckel, R.: Generic Finiteness for Dziobek Configurations. Trans. Amer. Math. Soc. 353, 4673--4686 (2001).
			
			\bibitem{Penrose}  Penrose, R.:. A generalized inverse for matrices. Math. Proc. Cambridge Philos. Soc. 51 (3), 406–13 (1955).
			
			
			
			\bibitem{Smale} Smale, S.: Topology and mechanics. II. The planar n-body problem. Invent. Math. 11, 45--64 (1970).
			
			
			\bibitem{SmaleII} Smale, S.: Mathematical problems for the next century. In Mathematics: Frontiers and Perspectives, pp. 271--294. Amer. Math. Soc., Providence, RI (2000).
			
			\bibitem{Williams} Williams, W. L.: Permanent configurations in the problem of five bodies. Trans. Amer. Math. Soc. 44, 563–579
			(1938).
			
			
			
			
		\end{thebibliography}
	\end{document}